# Handling Packet Dropouts and Random Delays for Unstable Delayed Processes in NCS by Optimal Tuning of $PI^\lambda D^\mu$ Controllers with Evolutionary Algorithms

Indranil Pan[a], Saptarshi Das[a,b], and Amitava Gupta[a,b]

a) Department of Power Engineering, Jadavpur University, Salt Lake Campus, LB-8, Sector 3, Kolkata-700098, India. Email: indranil.jj@student.iitd.ac.in, indranil@pe.jusl.ac.in

b) School of Nuclear Studies & Applications (SNSA), Jadavpur University, Salt Lake Campus, LB-8, Sector 3, Kolkata-700098, India. Email: saptarshi@pe.jusl.ac.in, amitg@pe.jusl.ac.in

**Abstract:**
The issues of stochastically varying network delays and packet dropouts in Networked Control System (NCS) applications have been simultaneously addressed by time domain optimal tuning of fractional order (FO) PID controllers. Different variants of evolutionary algorithms are used for the tuning process and their performances are compared. Also the effectiveness of the fractional order $PI^\lambda D^\mu$ controllers over their integer order counterparts is looked into. Two standard test bench plants with time delay and unstable poles which are encountered in process control applications are tuned with the proposed method to establish the validity of the tuning methodology. The proposed tuning methodology is independent of the specific choice of plant and is also applicable for less complicated systems. Thus it is useful in a wide variety of scenarios. The paper also shows the superiority of FOPID controllers over their conventional PID counterparts for NCS applications.

**1. Introduction:**

In the wake of development and cheap availability of viable communication systems, networked control systems have received increasing attention from a diverse background of research communities [1], [2]. As this field involves a fusion of multidisciplinary concepts, different research groups have tried to address the inherent issues in NCS from different viewpoints. Since in an NCS, the traditional control loop is closed over a real time communication network, various network constraints affect the performance of the control loop. In particular, delays introduced by the network (due to transmission, routing etc.) and packet dropouts (due to buffer overflows etc.) are of prime concern while assessing the control loop performance. These delays and packet dropouts not only degrade the performance, but can also make a control system unstable [3]. Alleviation of these problems may be achieved in a multitude of ways. One approach is to improve the various transmission protocols which are responsible for the routing of the packets in the network [4]. Others mostly look at improvising and implementing various control policies to take network delay and dropout into consideration [5]-[6]. In [7] the NCS is viewed as an asynchronous dynamical system with the network existing only between the sensor and the controller. The stability analysis of such systems is then done using Bilinear Matrix Inequalities in a Lyapunov framework. Other approaches have analyzed NCS using the concept of switched systems [5]-[6] and lifted sampling period [8]. The focus of these papers is mainly the stability aspect of the NCS in the presence of variable delay and packet dropout. However for process control applications, various other performance criterions like set point tracking, load disturbance rejection etc. are also of prime importance and are not addressed adequately by these literatures. Other literatures like [9]-[14] have tried to address these issues of process control in NCS applications by incorporating the concept of jitter margin for variable time delays and formulating various PID tuning rules for the same. However, in [9]-[11], [13]-[14] the NCS models do not account for packet drops and are limited to only variable delays which are unrealistic in a real time network. In the present paper both variable delays and packet drops encountered in a practical NCS application is incorporated in the tuning methodology and the major issues of process control application, like set-point tracking, load disturbance rejection etc. are addressed with the help of evolutionary algorithms. The tuning methodology is applicable to both PID and Fractional Order PID controllers and is illustrated by credible simulation studies.

Fractional order calculus has existed for over three centuries. These types of mathematical models of systems are capable of representing the natural phenomena in a more general way and do not approximate the processes by considering that the order of the governing differentials are integers only [15]-[16]. Recent hardware realizations of FO controllers [17]-[18] have brought renewed interest in such type of systems. The traditional notion of the PID controller has been extended to incorporate a more flexible and generic structure $PI^\lambda D^\mu$ by Podlubny [19], with the fractional differ-integrals as the design variables along with the controller gains. Several intelligent optimization algorithms have been used to tune these FOPID controllers for various objectives. Dominant pole placement based optimization problems have been used to design $PI^\lambda D^\mu$ controllers using Differential Evolution (DE) in Biswas *et al.* [20]. Optimization of a weighted sum of Integral of Absolute Error (IAE) and Integral of Squared Controller Output (ISCO) has been done to find out the controller parameters with Genetic Algorithm (GA) by Cao, Liang & Cao [21] and with Particle Swarm Optimization by Cao & Cao [22]. Other integral performance indices based FOPID controller tuning has been attempted by Zamani *et al.* [23] and Lee & Chang [24]. In general fractional order PID controllers give better results in meeting various stringent control system performances over their integer order counterparts due to their extra degree of flexibility i.e. the presence of 5 parameters of tuning as opposed to 3 in PID controllers. This improved performance is also the motivation of using fractional order controllers in the present paper.

Some contemporary researchers have also focused on the implementation of the FOPID controller in network applications. In [25], FO systems have been applied to a DC motor implemented over the network. They have considered only the time delay due to the networks and have analyzed the system considering the worst case constant time delay. However this method does not reflect the actual scenario, as a plant which gives a stable response when tuned with a worst case value of constant time delay, might be unstable in the presence of the same value of stochastic time delay as indicated by Hirai & Satoh [26]. Fractional order controllers have also proved better than normal PID controllers in the synchronization of networked motion control systems as in [27]. Delay dynamics in real communication network have been classified as an $\alpha$-stable process [28]-[30] and fractional order controllers have proved good at compensating such delays in closed loop systems [30]-[32]. FOPID controllers have also been tuned for large jitter margin, and have been tested using hardware in the loop simulations in Bhambhani *et al.* [29]. However in these literatures fractional order PID controllers have not been analyzed to incorporate classical process control objectives or presence of packet drops in the NCS and is investigated in the present paper. Other NCS applications with FO systems and FO controllers are also becoming popular like gain and order scheduling controller [33], distributed co-ordination of networked systems [34]-[35], remote stabilization [36], synchronization [37] etc.

Attempt has been made by contemporary researchers to handle the stochastic variation in network induced delays like Pan, Das & Gupta [38] with optimal fuzzy PID controller using three stochastic algorithms; Huang, Bai & Li [39] with discrete PID and self-tuning fuzzy PID controller; Zhang, Shi & Mehr [40] with $H_\infty$-infinity based PID controller with prescribed disturbance and noise attenuation; Bjorkbom & Johansson [41] with Internal Model Control (IMC) based PID controller, Sala, Cuenca & Salt [42] and Cuenca *et al.* [43] with multi-rate controllers etc. This paper gives considerable improvement over existing literatures by treating the model as a hybrid system (discrete network model and continuous time plant model), handling the issues of both delay and packet dropout simultaneously by tuning with various evolutionary algorithms and also implementing buffers to filter the out of order packets. In this paper, it is also attempted to analyze the effectiveness of the fractional order controllers in handling the network delays and packet dropouts in an NCS application. Specifically, given a bounded, stochastically varying network delay and packet dropout rate, we tune the PID and $PI^\lambda D^\mu$ controllers with various evolutionary algorithms to get an optimal performance. The controllers are tested for load disturbance rejection and their objective functions compared to analyze the effectiveness of the different standard tuning algorithms and also the controllers themselves. The evolutionary algorithms used are standard algorithms which are widely used by the research community for a multitude of applications. Since the focus of the paper is on controller tuning, hence no modification over the existing algorithm is done and the performances of these algorithms are compared for this specific application. This is also advantageous from the control system engineer's view point since no additional customization of the algorithms need to be made and can be implemented right away [44]. The controller tuning is offline. The maximum upper bound of the delays and the percentage of packet losses need to be known in advance and the controller can be tuned with GA and DE to obtain controller parameters. Since the controller parameters are static and the tuning is offline, the issue of quick convergence in real time does not arise.

The rest of the paper is organized as follows. Section 2 discusses about the networked FOPID controller tuning scheme for two classes of unstable processes, considering stochastically varying objective function. Section 3 highlights the different evolutionary algorithms used for controller tuning. Section 4 shows the effectiveness of the proposed tuning methodology for closed loop control of unstable processes over network in terms of control performance and control signal. Validation with respect to other less complicated processes, delay distributions etc. have also been shown. The paper ends with the conclusions in Section 5, followed by references.

## 2. Control over communication network and scope for stochastic optimization based controller tuning:
### 2.1. Networked FOPID control scheme:

A NCS with a general PID/FOPID controller in the loop may be represented as in Fig. 1. The FOPID controller shown in Fig. 1 reduces to a simple PID controller when the integro-differential orders of the controller are unity. In Fig. 1, $\tau^{SC}$ represents the sensor to controller delay in the feedback path and $\tau^{CA}$ represents the controller to actuator delay in the forward path. The buffers are implemented at the receiving ends of the network to filter the out-of-order packet arrival. The network additionally drops packets with a certain probability, i.e. the packets are sent from the sending end of the network and eat up the bandwidth of the network, but these are never delivered to the destination node.

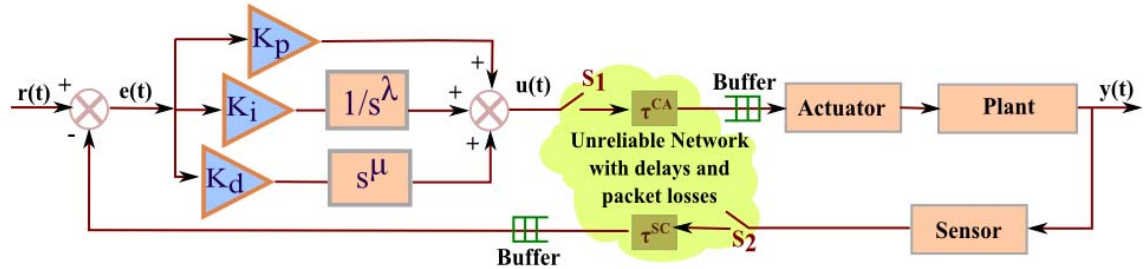

Fig. 1. Schematic diagram of the NCS augmented with an optimal FOPID controller to handle packet drop-outs and random delays.

The sensor is time driven and samples the process output at fixed instants of sampling time. The sensor sends the packets to the controller along with a time stamp. The PID/FOPID controller is also time driven. Generally the convention is to use an event driven controller which calculates the control signal as soon as it is received and sends it over the network to the actuator. This eliminates the additional time till the next sampling instant that the controller must wait before it transmits the control signal as in the time driven case. But in safety critical applications event driven controllers are not used since specific cases may occur where after a long delay, many control signals come together within the same sampling time thus increasing the instantaneous network load to a high value which is undesirable. Also, implementation of event-driven controllers is difficult in actual hardware [45]. Hence the choice of this type of controller is made in the present paper.

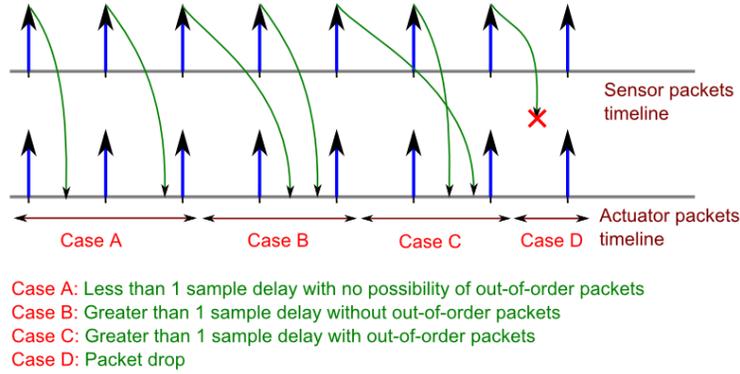

Case A: Less than 1 sample delay with no possibility of out-of-order packets
Case B: Greater than 1 sample delay without out-of-order packets
Case C: Greater than 1 sample delay with out-of-order packets
Case D: Packet drop

Fig. 2. Timing diagram of the NCS explaining occurrence of out-of-order packets.

Fig. 2 shows the timeline of the packets at the sensor and those at the actuator buffer with representative cases encompassing the various scenarios that might be encountered. In Case A the total time delay from the sensor to the actuator buffer is less than 1 sample time, hence there is no scope for out of order packets. In Cases B and C the maximum upper limit of the delays encountered is greater than one sample time. In some situations as in Case C, out of order packets may arrive which need to be discarded to prevent performance deterioration. In Case D a packet may start from the sensor but may never reach the actuator. This case might arise due to overflows in the buffers of the network queues, bit errors in transmission etc. Due to these variable delays and packet drops at arbitrary instants of time, the controlled process output and the control signal becomes stochastic in nature. The approach followed in the present work is to formulate an objective function which is minimized to ensure good control performance with a stochastic plant input. The optimization function and the optimization technique adopted are presented in the next subsections.

### *2.2. Time domain integral performance index based tuning of networked process controllers via stochastic optimization for randomly-varying objective function:*

The integral performance index ($J$) chosen for optimal PID/FOPID controller tuning is the summation of the Integral of Time multiplied Absolute Error (ITAE) and the Integral of the Squared Controller Output (ISCO). i.e.

$$J = \int_0^\infty \left[ w_1 \cdot t \left| e(t) \right| + w_2 \cdot u^2(t) \right] dt = \left( w_1 \times ITAE \right) + \left( w_2 \times ISCO \right) \qquad (1)$$

It is worth mentioning that the weights $w_1$ and $w_2$ have been introduced in the objective function (1) with a provision of balancing the impact of the error and control signal. In the present simulation study we have considered equal weights for the two objectives to be met by the controller as such the minimization of the error index and control signal is equally important.

Equation (1) is a time domain optimality criterion which ensures faster settling time by penalizing errors to a greater extent at the later stages (due to multiplication of the time term) and also reduces the peak overshoot (due to the multiplication of the absolute error term). The controller output can also become very large and result in actuator saturation and integral wind-up, hence it is also minimized by including it in the performance index. Other performance indices like ITSE (Integral of Time multiplied

Squared Error), ISTES (Integral of Squared Time multiplied Error Squared) etc. with higher powers of time and error may be considered instead of ITAE. The higher powers in time and error penalizes the output more at later stages and hence gives very fast rise and settling time. But for a sudden change in set-point this kind of strict criteria gives very high value of controller output which might result in actuator saturation and therefore integral wind-up. Hence ITAE is considered as the error index for optimally finding the controller parameters that ensures efficient set-point tracking.

In the presence of stochastically varying delays and packet-drops in the network the objective function (1) will have different values for the same set of controller parameters, depending on the time instants of packet arrival. This can be intuitively understood from the following logic. The cases A, B, C and D in Fig. 2 occur randomly without any pre-specified sequence. Hence for the same set of plant parameters and controller gains, the process output would evolve through a different trajectory if evaluated multiple times. This would result in different values of the objective function $J$ in (1) since it involves the integral of the plant and controller output. A representative function showing the nature of such stochastic variation is shown in Fig. 3. In Fig. 3 the landscape of the objective function varies by a small amount at different instants of time but the variation is small and hence the overall nature of the landscape is preserved (i.e. the minima and maxima do not shift positions appreciably). Hence, the objective function for time domain optimal controller tuning is stochastic in nature and must be handled by intelligent non-gradient based algorithms [44].

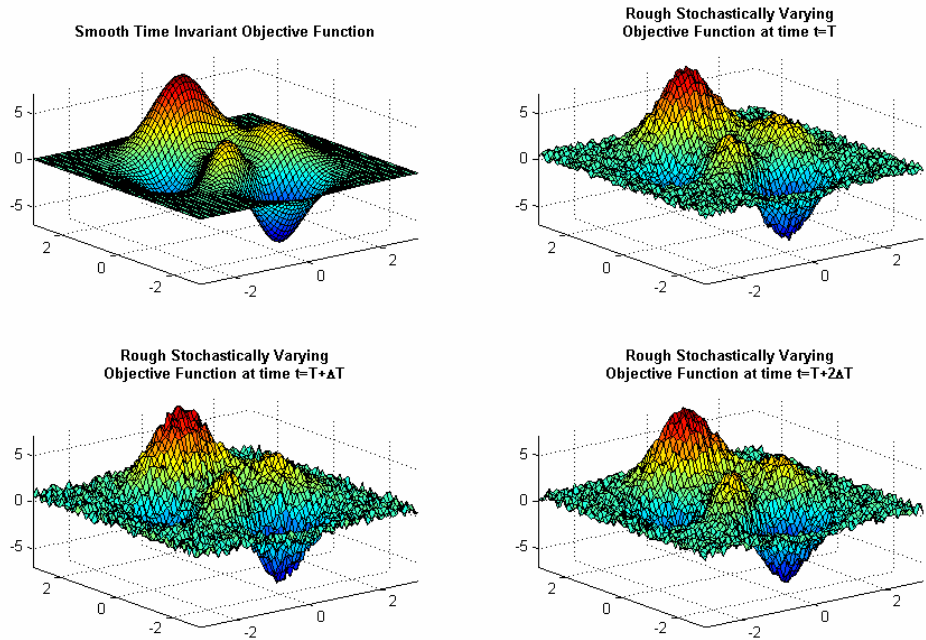

Fig. 3. Example of a stochastically varying objective function and concept of its minima.

Fig. 3 shows the difference between a smooth time invariant representative objective function and its corresponding rough, stochastically varying counterpart. Various gradient based algorithms like Nelder-Mead Simplex algorithms and others exist for finding the minima of smooth differentiable functions but they fail to work for rough and stochastically varying objective functions as in the present case. It is interesting to note however that inspite of the stochasticity, the overall functional landscape does not change appreciably and the minima or maxima shifts by only very small amounts. Thus a stochastic optimization method like GA or DE is suitable in this case to find out near optimal solutions to the problem in the presence of random variation in the objective function [44]. The original GA and DE variants are modified and the values of the stochastic fitness function are evaluated multiple times for the same inputs and are then averaged to obtain the final output during each evaluation. This essentially is a form of calculation of the statistical expectation of the randomly varying objective function at each function evaluation.

### *2.3. Fractional Order PID controller in NCS applications:*

Till date, the classical PID controllers dominate the process industries. The concept of fractional order PID controller comes from the availability of two extra tuning knobs viz. the differ-integral orders to meets additional design specifications. Recent research shows that fractional order $PI^\lambda D^\mu$ controllers are capable of producing finer control system design over the conventional PID controllers [15]-[17]. $PI^\lambda D^\mu$ controller tuning for deterministic systems have been attempted by contemporary researchers like Valerio & Sa da Costa [46], Monje *et al.* [47], Padula & Visioli [48], Das *et al.* [49], Kakhki & Haeri [50] etc. However, tuning FOPID controllers for stochastic systems i.e. processes over network with variable delays and packet-drops have not been attempted yet and is the focus of the present paper.

The FOPID controller considered here for optimal time domain tuning, is represented in non-interacting or parallel structure. The transfer function of the FOPID controller is given as:

$$C^{FOPID}(s) = K_p + \frac{K_i}{s^\lambda} + K_d s^\mu \qquad (2)$$

Equation (2) is actually a generic form with higher degrees of freedom for tuning and reduces to the integer order PID by simply setting the values of $\{\lambda, \mu\} = 1$. The fractional order differ-integrals $\{\lambda, \mu\}$ are basically infinite dimensional linear filters. However band-limited realization of $PI^\lambda D^\mu$ controllers is necessary for its hardware implementation. In the present simulation study each fractional order element has been rationalized with Oustaloup's recursive filter [51] given by the following equation (3)-(4). If it be assumed that the expected fitting range or frequency range of controller operation is $(\omega_b, \omega_h)$, then the higher order filter which approximates the FO element $s^\gamma$ can be written as:

$$G_f(s) = s^\gamma = K \prod_{k=-N}^{N} \frac{s + \omega'_k}{s + \omega_k} \qquad (3)$$

where the poles, zeros, and gain of the filter can be evaluated as:

$$\omega_k = \omega_b \left(\frac{\omega_h}{\omega_b}\right)^{\frac{k+N+\frac{1}{2}(1+\gamma)}{2N+1}}, \omega'_k = \omega_b \left(\frac{\omega_h}{\omega_b}\right)^{\frac{k+N+\frac{1}{2}(1-\gamma)}{2N+1}}, K = \omega_h^\gamma \qquad (4)$$

In (3) and (4), $\gamma$ is the order of the differ-integration and $(2N+1)$ is the order of the filter. Present study considers a 5$^{th}$ order Oustaloup's rational approximation [15] for the FO elements within the frequency range $\omega \in \{10^{-2}, 10^2\}$ rad/sec which is most common in process control applications.

The main focus of the present paper is to study performance of complicated processes e.g. unstable processes which are hard to control with PIDs over network with an optimally tuned FOPID controller. Contemporary researchers like Bhambhani *et al.* [29] uses maximization of jitter margin as one of its objectives and the tuning rules used demonstrate that a FOPI controller produces a larger jitter margin for lag dominated First Order Plus Time Delay (FOPTD) process controlled over a network while a PID controller produces a larger jitter margin for a delay dominated plant. Thus, to show the wide applicability, the proposed methodology has also been validated for two relatively lesser complicated FOPTD processes like a delay dominated and a lag-dominated one, as representative cases.

### *2.4. Handling unstable processes over NCS:*

Two representative plants have been considered for control performance analysis in NCS application with the proposed methodology. Huang & Chen [52] introduced two class of open loop unstable processes which are most common in process control viz.

(a) First Order Delayed Unstable Process (FODUP):
The structure of such a process can be described as

$$P_{FODUP}(s) = \frac{Ke^{-Ls}}{(Ts-1)} \tag{5}$$

(b) Second Order Delayed Unstable Process (SODUP):
The structure of these class of process are described as

$$P_{SODUP}(s) = \frac{Ke^{-Ls}}{(T_1 s - 1)(T_2 s + 1)} \tag{6}$$

In (5) and (6), the system parameters $\{K, L, T\}$ represent the system's dc-gain, transport-delay and time-constant respectively. The stability of such an open loop unstable process is highly sensitive compared to other stable sluggish processes and thus the performance degradation and the true potential of the proposed design methodology can be effectively compared with other conventional methods. Also, the presence of time delay along with the unstable pole makes the system more difficult to control and even more so with stochastic network delays and packet dropouts.

For the sake of simulation study an FODUP type process is considered as studied by Visioli [53].

$$P_1(s) = \frac{e^{-0.2s}}{(s-1)} \tag{7}$$

The FODUP Process $P_1$ has a single unstable pole and the time constant is much larger than the delay. Next a higher order unstable process [54] has been considered in a NCS test-bed given by,

$$\tilde{P}_2(s) = \frac{e^{-0.5s}}{(5s-1)(2s+1)(0.5s+1)} \tag{8}$$

Such higher order open loop unstable process has been reduced to standard SODUP template (6) in [52] as the following transfer function:

$$P_2(s) = \frac{e^{-0.939s}}{(5s-1)(2.07s+1)} \tag{9}$$

Performance of these two unstable processes (7) and (9) are tested with an optimally tuned PID/FOPID controller with the consideration of randomness in network induced delay and packet drop-out in the following subsections.

### 3. Evolutionary algorithms based PID/FOPID controller tuning:

Intelligent stochastic algorithms have been widely applied to controller tuning as in [20]-[24]. Since the closed loop system is stochastic in nature due to randomness in network induced delays and packet drop-out probabilities, the objective function (9) varies stochastically and hence conventional gradient based algorithms for optimization fail to give acceptable solutions. Evolutionary algorithms do not need any gradient information and have proved expedient in such type of noisy and uncertain environments [55]. Two evolutionary algorithms have been explored for optimal tuning of FOPID and PID controllers viz. Genetic Algorithm and Differential Evolution along with few of its variants. Both these evolutionary algorithms minimize the objective function (1) and have the decision variables as $\{K_p, K_i, K_d\} \in [0,100]$ and $\{\lambda, \mu\} \in [0,2]$. The details of the evolutionary algorithms are briefly introduced in the following sub-sections.

#### *3.1. Genetic Algorithm (GA):*

GA is a stochastic optimization procedure, based on the process of natural biological evolution. Here each solution vector is considered to be a genome which undergoes reproduction, mutation and crossover to yield fitter genomes in the next generation. The fitness is usually calculated by an objective function for each of the genomes in each generation.

The solution vectors with a higher value of fitness function can undergo reproduction to create more copies of themselves in the next generation, based on probabilistic decisions. In crossover, a certain percentage of the population exchange information among themselves to give rise to fitter solution vectors. In Mutation, a randomly selected portion is sometimes altered to yield fitter individuals. Thus these three processes go on concurrently in each generation until a pre-specified number of generations are attained or the function value does not change over consecutive generations. Usually for most optimization problems, the probability of mutation is kept small and that of crossover is kept large. In this case the mutation fraction is 0.2 and the crossover fraction is 0.8.

The population size must also be chosen judiciously, as making the population size huge would require more number of function evaluations and would increase computational time. Also too few individuals might lead to premature convergence. Generally as a thumb rule the number of individuals must be at least greater than the number of dimensions of the problem. In our case we have chosen the population size as 20 to ensure a balance between both. Also another parameter called the elite count is used, which represents the number of fittest individuals in each generation who would definitely go into the next generation. If this value is too big in comparison to the population size, then it would lead to a dominance of the initially obtained fitter individuals, which would restrict the exploration of the solution vectors in the search space. In our case we have used a value of 2 for the elite count.

Various improved methodologies have been proposed over the standard Genetic Algorithm as in [55]. However using conventional GA is also an option in these noisy environments as GA has self averaging nature [56] i.e. the solution vectors having good fitness values propagate through the generations and survive. Convergence with the standard GA takes more time, which can be reduced by exploiting the parallel structure of the algorithm [57]. In each generation, the objective function can be evaluated for each of the individuals in parallel since these evaluations are not dependent on the evaluation of other individuals in the current generation. In the present case the objective function has been modified so that it can be evaluated multiple times for the same set of inputs and is averaged to give the expected value of the function.

*3.2. Differential Evolution (DE):*

DE basically emerged as a more refined version of the GA with subtle changes to overcome some of the disadvantages of the GA. Like the GA, the DE initializes a random population of solution vectors from the solution domain. However unlike the GA where the solution vector is encoded as a bit representation, the DE uses floating point numbers to represent the solution vectors. Thus compared to the "bit flipping" approach with logical operators (like XOR) in GA, the DE performs the mutation and crossover with arithmetic operators, which lowers computational complexity and facilitates greater flexibility in the design of the mutation distribution.

The classic DE algorithm (DE/rand/1) randomly chooses three solution vectors from the initial population and adds a weighted difference of the first two vectors to the third vector (base vector). This process is called mutation and is repeated until an intermediate mutant population, equal in number to the original population is created. Each of these mutant vectors undergoes a crossover with a solution vector (known as target vector) from the initial population to generate another set of trial vectors. The crossover probability ($Cr \in [0,1]$) specified by the user, dictates the fraction of parameter values that are taken from the mutant vector. Finally the trial vector is selected as a solution vector in the next generation if its objective function value is lower than that of the target vector. Otherwise the target vector is selected to go into the next generation. Thus this is an elitist selection as the current best vector of the population can be replaced only by a fitter vector. This process is iterated and the program terminates when a specified number of iterations are exceeded or the value of the objective function falls below a pre-specified level.

In this simulation different variants of DE are adopted for comparison of these heuristic methods. DE has been shown to give satisfactory results for noisy real world optimization problems as in [58] without the need for rigorous parametric study. The parameters chosen in the simulation are the ones which have shown to work for a wide variety of optimization problems as reported in [59]. A parametric study of these algorithmic parameters are possible to speed up the convergence and get a slight improvement in accuracy, but it entails rigorous computer simulation and is beyond the scope of the present work. For all of the DE variants the number of particles is 20 and the maximum number of generations is 200. The variants of DE mainly differ in the way that the mutation operation is done.

*3.2.1. DE/rand/1 (classical DE):*

The nomenclature rand implies that the base vector in DE is randomly chosen, and 1 implies that 1 vector difference is added to it. This is the original classical version of the DE algorithm. The difference vector is multiplied by a factor $F$ to get the weighted version of the same. The mutant parameter vector $v_i$ is thus generated as

$$v_i = x_{r_0} + F(x_{r_1} - x_{r_2}) \tag{10}$$

where, $x_{r_0}, x_{r_1}, x_{r_2}$ are the three randomly chosen population members. Generally $F$ can vary between [0-2] and the DE algorithm is somewhat sensitive to the choice of $F$. As indicated in [59] a good initial guess is between [0.5-1]. In this case we have chosen $F$ as 0.85. The crossover probability is taken as 0.5 and helps maintain diversity of the population.

*3.2.2. DE/local-to-best/1:*

In this case instead of equation (1) the mutation expression is given by

$$v_{i,g} = old_{i,g} + (best_g - old_{i,g}) + x_{r_0} + F(x_{r_1,g} - x_{r_2,g}) \tag{11}$$

where, $old_{i,g}$ and $best_g$ are the $i^{th}$ member and the best member respectively of the previous population. This version of DE tries to maintain a balance between robustness and fast convergence [59].

*3.2.3. DE/best/1 with jitter:*

In this variant the base vector is not randomly chosen and is the one with the lowest objective function in the present generation. Compared to the random vector selection, this version generally speeds up the convergence and reduces chances of stagnation but it lowers the probability of success [59]. To compensate for the loss of diversity due to this strategy, the scale factor $F$ (the weight with which the difference of two vector is multiplied while creating the mutant population), instead of being constant, is assumed to be a normally distributed random variable for each component of the solution vector. The DE mutation expression is given by

$$v_{i,g} = best_g + jitter + F(x_{r_1,g} - x_{r_2,g}) \tag{12}$$

where, $jitter = 0.0001 \times rand + F$

Since this jitter process multiplies each of the component of the difference vector with different values, it not only introduces a change in scale but also results in a change in orientation of the resultant vector. This makes it fundamentally different from the classic DE where $F$ is a constant. This variant of DE is mainly used for faster convergence in problems which have a small dimensionality and a reasonably small population size.

*3.2.4. DE/rand/1 with per vector dither:*

In this variant the scale factor is chosen anew, from a normally distributed random variable, for the whole vector instead of each component. This is known as "per vector dithering". The mutation equation is given by

$$v_{i,g} = x_{r_{0,g}} + dither \times (x_{r_{1,g}} - x_{r_{2,g}}) \qquad (13)$$

where, $dither = F + rand \times (1-F)$

This makes the solution algorithm more robust and removes the need for carefully tailoring the value of $F$ for each particular type of problem.

*3.2.5. DE/rand/1 with per generation dither:*

Here the dither is applied to the whole generation instead of each vector in the generation. This also increases the robustness of the algorithm [59]. In all of the above algorithms a large penalty function has also been incorporated in the optimization process for very large value of $J$ (1) to avoid parameter search with unstable closed loop response as suggested by Zamani *et al.* [23]. Also as in the case of GA the objective function for DE has been modified so that it can be evaluated multiple times for the same set of inputs and is averaged to give the expected value of the function.

**4. Simulations and Results:**

*4.1. MATLAB based simulation study of a NCS with packet dropout and variable delay:*

As mentioned before, the whole control loop over a communication network is actually a hybrid system, in the sense that both continuous order plant dynamics coexist with the discrete packet level network dynamics together in the same model. Hence modeling this kind of a system needs to capture the simultaneous effects of both of these dynamics taken together. In this case MATLAB Simulink [60] is used for the continuous time plant and SimEvents [61] is used to add the discrete network in the loop. For all simulations, the sampling time is assumed to be 0.01 seconds. A Time Stamp Order (TSO) processing is implemented at both the receivers viz. the controller and the actuator to keep out of sequence packets ordering. Thus out of sequence packets are dropped. The simulator allows the user to specify the maximum delay that a packet can face and also the maximum percentage dropout assuming uniform distribution. The characteristics of the network, i.e. the probability of packet dropout and the lower and upper bounds of stochastic delay with the delay distribution can be specified by the user. When the delay of each packet is less than 0.01 seconds then the sequence of the packets are maintained, but when the delays are greater than one sampling period, then the packets arrive out of sequence. This was found to be very detrimental to the performance of the control system. So a mechanism is incorporated whereby the packets carrying the data are time stamped and there is a buffer at each of the receiving ends of the network, which compares the time stamp with the last arrived packet and lets the packet to pass through only if it is newer than the former. Thus out of sequence packets are dropped due to this which is in addition to the pre-specified network packet dropout level. Thus it is evident that greater the upper bound of the losses, the more is the chance of having out of order packets and hence more number of packets would be dropped at the buffer. This is also a very realistic scenario from the actual network viewpoint.

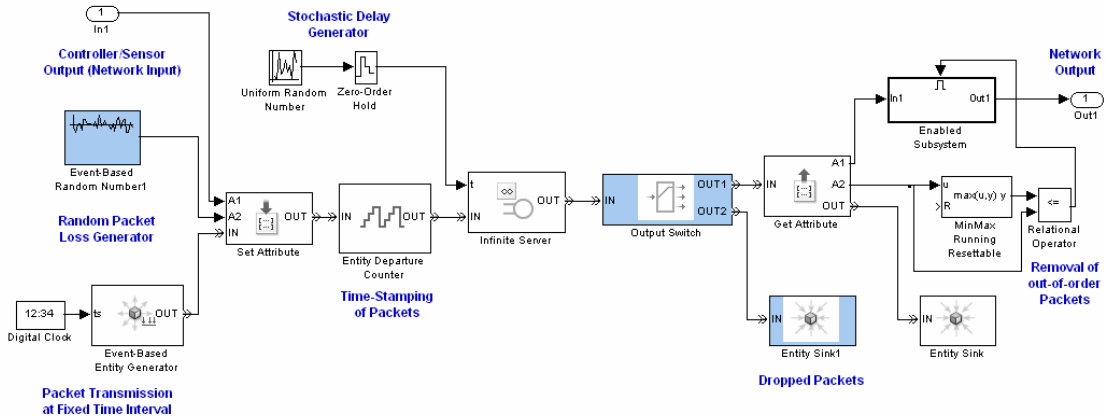

Fig. 4. Simulink and SimEvent based implementation of the NCS test-bed.

Fig. 4 represents the SimEvent based packet level implementation of the network model. The network packets are generated at fixed intervals of 0.01 seconds by the Entity Generator block. A random packet loss generator embeds a number 1 or 2 depending upon the probability of packet drop. The value from the controller or sensor is also embedded in the packet using the Set Attribute block. The packets are time stamped by assigning the packet number using the Entity Departure Counter. The Infinite Server block delays each packet by a random amount depending on the value from the Stochastic Delay Generator block. The Output Switch looks at the value of the packet loss attribute embedded in the packet and pushes it through the first output port if the value is 1 and through the second output port if the value is 2. The packets which go out through port 2 never reach the destination and are dropped. The packets coming out of output port 1 are compared to see if their time stamp is newer than the last received packet and are passed to the destination of the network resulting in a successful transmission. The simulation setup presented in this sub-section is used to study the performance degradation of well tuned control loops under packet drop out and variable delay situations and also optimum controller PID/FOPID tuning with these irregularities in the loop.

### *4.2. Performance degradation of well tuned control loops with stochastic consideration of the network delay:*

For comparison, the network delays ($\tau^{SC}$ and $\tau^{CA}$) are lumped with the process delay and tuned with optimal time domain method with a PID controller. Lumped static delays are more common in process control and are much easier to handle. Hirai & Satoh [26] have shown that same amount of stochastic delay may lead to instability when the lumped delay gives a stable response. With this motivation an optimal PID controller is first tuned with a lumped static delay approximation and with no network delay consideration (Table 1). Then the random delay in the closed loop is increased to show that the performance degradation of the time response curves (Fig. 5) if the randomness is not considered within the optimization process for controller tuning which is the main focus of this paper.

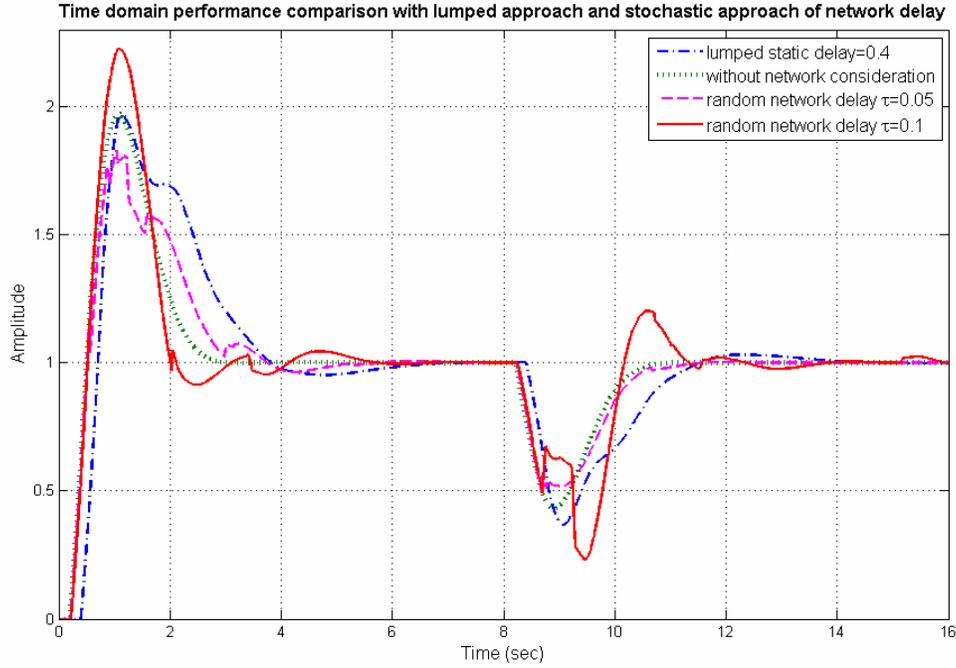

Fig. 5. Time response of optimal PID tuning for $P_1$ with lumped approach and stochastic approach of network delay.

Table 1: GA based tuning of PID controllers for $P_1$ at different network conditions

| Network Condition | $J_{min}$ | $K_p$ | $K_i$ | $K_d$ |
|---|---|---|---|---|
| without network | 49.20556694 | 2.688684627 | 1.486143944 | 0.045784858 |
| static delay | 52.68181486 | 2.601225912 | 1.203599975 | 0.57165169 |

As is evident from Fig. 5, a small amount of stochastically varying network delay is much more detrimental to the system performance than a greater amount of static delay. Progressive increase of the stochastic delay leads to instability of the plant much before the upper limit of stochastic delay attains the value of the constant lumped delay.

### *4.3. Performance degradation of well tuned control loops due to out of order packets and handling packet drop out with buffers:*

When the network delays are greater than one sampling time, each packet may suffer a different amount of stochastic delay due to different routing paths, queuing at the buffers etc. and thus in certain cases the packets which was sent earlier from the sending node of the network is received after the arrival of newer packets (Fig. 2). Fig. 6 shows the detrimental effect in the closed loop response when the out of order packets are allowed to remain in the loop. The packets suffer a stochastic delay varying between 0 to 0.05 seconds and a probability of 5% packet dropout in both the forward and the feedback paths. A time stamped buffer is implemented after the receiving node of each network path, which only allows the newer packets to pass through and discards the out of order packets. The system response with this mechanism incorporated is also shown in Fig. 6 which shows significant performance improvement over the previous approach. In our optimization procedure we adopted a similar mechanism for the network with buffers implemented in the forward and feedback paths (Fig. 1).

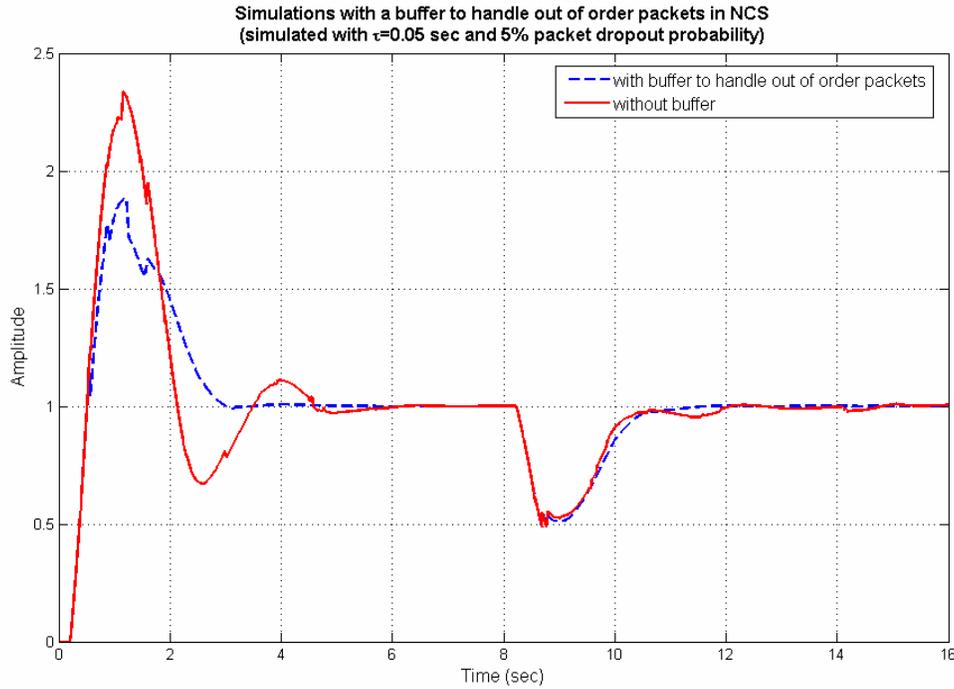

Fig. 6. Simulations with a buffer to handle out of order packets for $P_1$ over NCS.

### *4.4. Optimal PID and FOPID controller tuning for unstable processes with the consideration of randomness in network induced delays and packet dropouts:*

A PID and a $PI^\lambda D^\mu$ controller are now tuned by minimizing the objective function (1) using different evolutionary algorithms discussed in Section 3. The optimal integer and fractional order controllers have been tuned considering 0.1 sec random delay both in the forward and backward path of the NCS (Fig. 1) i.e. $\tau^{SC} = \tau^{CA} = 0.1\,\text{sec}$ and with a probability of packet drop-out to be 0.1 for both the FODUP and SODUP processes. The performance of the test FODUP process (7) has been compared on the basis of optimal tracking for a unit change in set point as well as suppression of unit load disturbance as in [53]. Table 2 shows the PID controller parameters along with the minimized value of the objective function (1) for different evolutionary algorithms. The value of $J_{\min}$ is actually the expected value since the objective function is stochastic in nature in the presence of variable delay and packet drops. This is followed by Fig. 7 that shows the time domain performance of the PID controller for unit set point change and load disturbance. Fig. 8 shows the PID controller output for these cases. Table 2 and Fig. 7-8 show that the DE variants are better than GA and are able to locate lower values of the objective function. PID controllers tuned with DE also give a better time domain performance especially in terms of peak overshoot. The load disturbance rejection for all the cases is nearly similar. However as is evident from the controller output in Fig. 8, the delays and packet drops result in a very oscillatory controller output with high amplitude in certain instants. As already discussed, to ensure that a large control signal does not saturate the actuator, the control signal $u(t)$ is also minimized as a part of the objective function (1).

It is worth mentioning that the random network delays and packet dropouts would lead to such a variation of controller output as in [38]. Some prediction mechanisms based on the Quality of Service (QoS) of the network can ameliorate the adverse effects to a certain degree. Other avenues might include designing of effective transmission protocols so that the random delays are handled adequately by the network protocol itself. Also, the ISCO in the performance index effectively reduces the band of oscillation in the control signal. With only a simple error minimizing criteria like [62], [48], [49] the band of oscillation in control signal would have been more. This is a particular problem introduced by the stochastic characteristic of the communication network. Since the primary aim of the controller is to maintain time domain optimality which is evident from the time responses, the control signal suffers to some extent (as it has been optimized by the tuning algorithm and not been allowed to be arbitrarily high) so as to suppress the effect of random delay in the time responses and yield a smooth closed loop dynamics.

Table 2: Tuning results of PID controller for FODUP process $P_1$

| Algorithm | $J_{min}$ | $K_p$ | $K_i$ | $K_d$ |
|---|---|---|---|---|
| DE/rand/1 | 49.61417 | 2.598099 | 1.329863 | 0.049072 |
| DE/local-to-best/1 | 48.98632 | 2.539656 | 1.248308 | 0.029662 |
| DE/best/1 with jitter | 49.03776 | 2.856348 | 1.800701 | 0.031918 |
| DE/rand/1 with per-vector-dither | 48.74167 | 2.725339 | 1.624656 | 0.026691 |
| DE/rand/1 with per-generation-dither | 49.59382 | 2.589976 | 1.283022 | 0.029582 |
| GA | 50.78769 | 2.511934 | 1.162209 | 0.010378 |

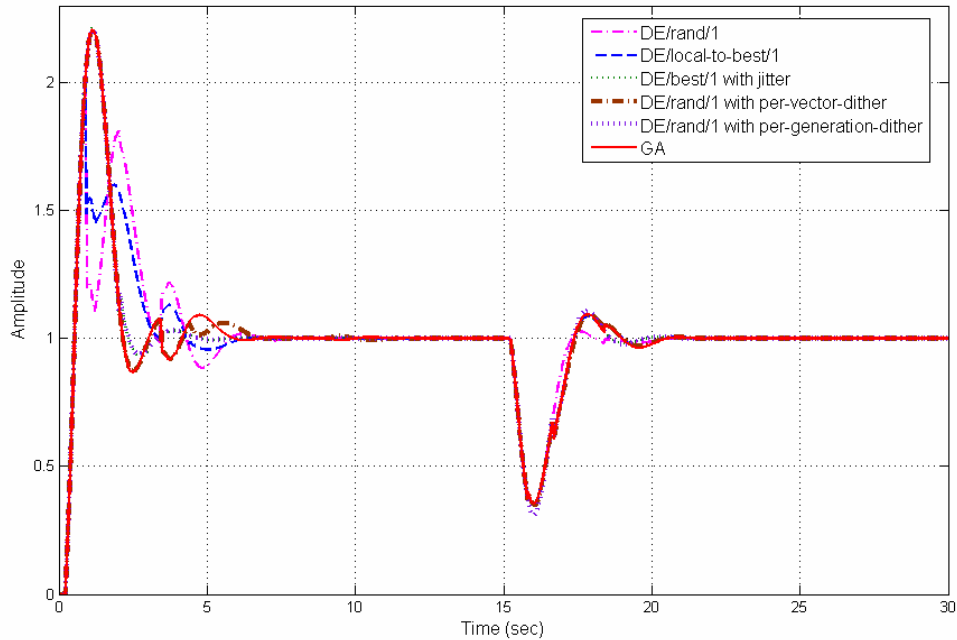

Fig. 7. Time response of $P_1$ over NCS with optimal PID controller.

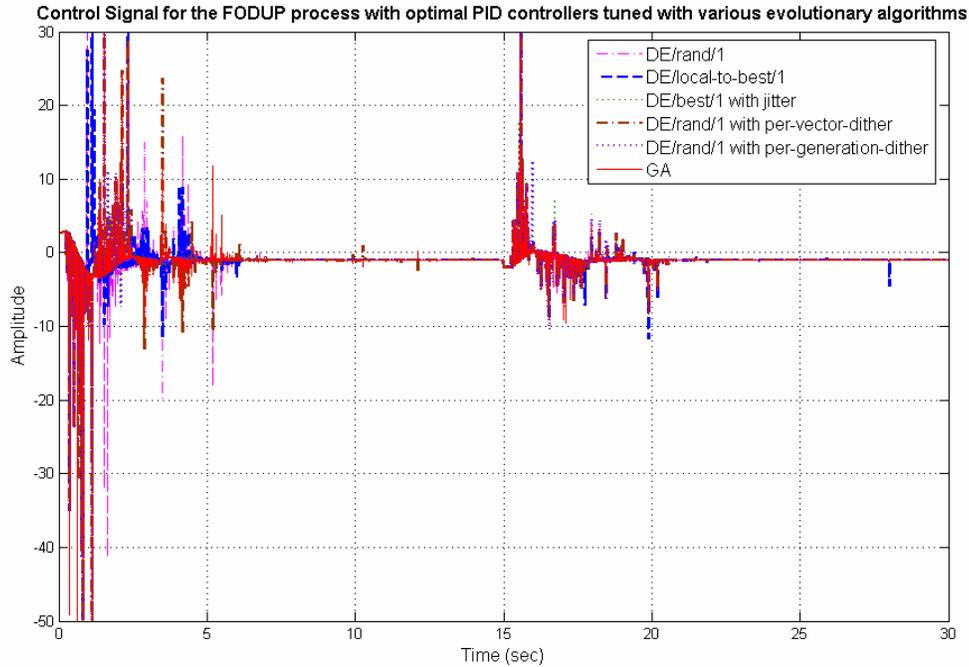

Fig. 8. Control signal of optimal PID controller for $P_1$ over NCS.

A well tuned PID controller corresponding to the minimum value of $J_{min}$ in Table 2 (i.e. with the algorithm DE/rand/1 with per-vector-dither) is now tested with varying differ-integral orders ($\lambda$ and $\mu$) as studied by Chen [63]. The corresponding ITAE, ISCO and their weighted sum ($J$) in Fig. 9-11 show a stochastic control surface, which can not be optimized with conventional gradient based optimization algorithms. That is why to minimize a control objective in such a rough stochastically varying environment evolutionary algorithms are employed.

In fact, Fig. 9-11, show the control surface for $\{\lambda, \mu\}$ variation for an initially tuned PID control loop. An improved performance can be expected if the five parameters of a $PI^\lambda D^\mu$ controller are tuned to meet the control objective (1) instead of optimally finding $\{\lambda, \mu\}$ for a stable PID control loop over NCS as studied by Chen [63]. Fig. 9-11 show the control performance is heavily dependent on the derivative order ($\mu$) than the integral order ($\lambda$). But simultaneous tuning of five parameters of a $PI^\lambda D^\mu$ controller may not yield similar dependencies on $\{\lambda, \mu\}$.

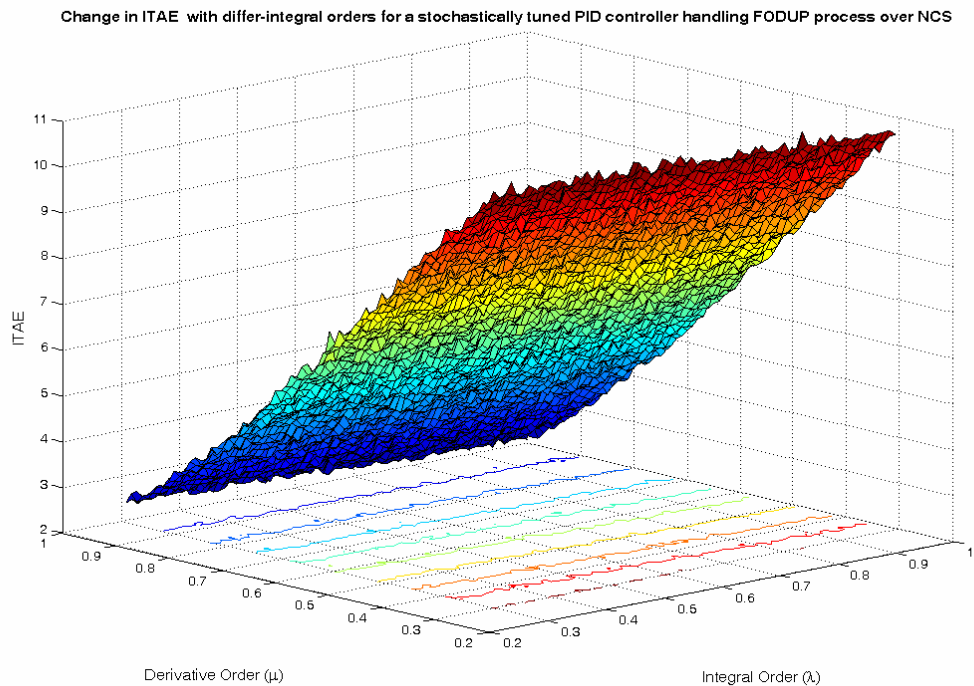

Fig. 9. Change in ITAE with differ-integral orders for a stochastically tuned PID controller handling FODUP process over NCS.

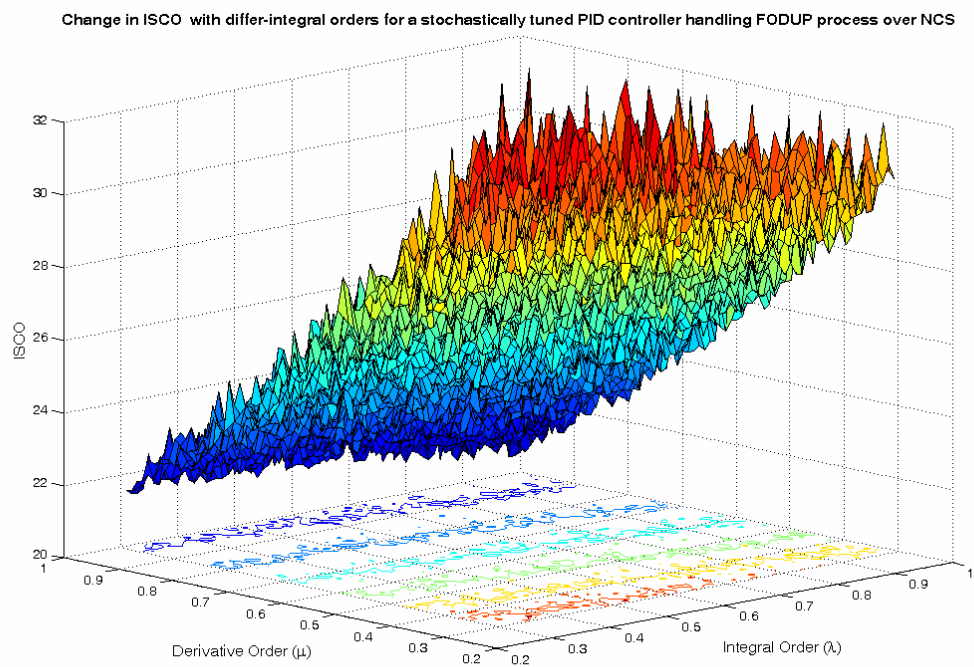

Fig. 10. Change in ISCO with differ-integral orders for a stochastically tuned PID controller handling FODUP process over NCS.

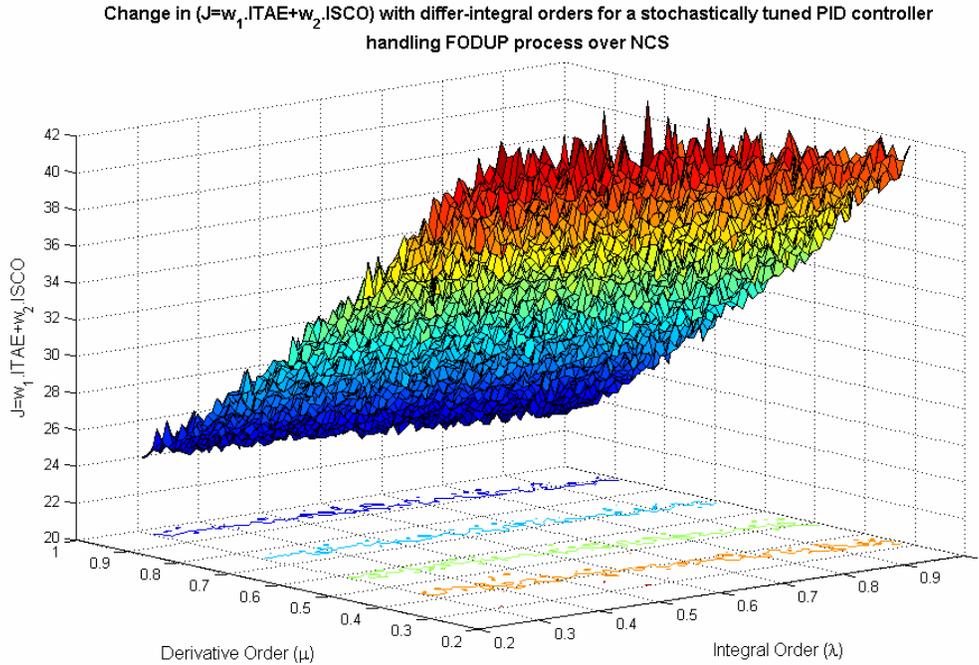

Fig. 11. Change in $J$ with differ-integral orders for a stochastically tuned PID controller handling FODUP process over NCS.

Table 3 shows the tuned values of the $PI^\lambda D^\mu$ controller parameters along with the expected minima of the objective function ($J_{\min}$) for the different evolutionary algorithms. The DE algorithms are slightly better than GA in these cases and the time domain performance of the controllers in terms of peak overshoot, settling time and load disturbance rejection are almost same. However the peak overshoot of the FOPID controllers are much lesser than their PID counterparts and the settling time is also faster incase of the FOPID controllers (Fig. 12). Fig. 13 shows the controller output for the FOPID controller and the abrupt changes in the control signal for the delay and packet dropout is significantly smaller (almost by a factor of 6) than the corresponding PID controller counterpart. Thus the actuator design need not be large and also this reduces the chances of actuator saturation. Thus optimally tuned $PI^\lambda D^\mu$ controllers have a significant advantage over conventional PID controllers when used for NCS applications as it has a higher capability of suppressing the adverse effects of random phenomena in a closed loop control system.

The optimal controller performances are reported next for SODUP process $P_2$ while minimizing the objective function (1). In Table 4, the minimum cost function corresponds to the finite time horizon of 40 seconds. For the heavily unstable process $P_2$, no stabilizing PID controller gains are obtained that can handle such significant amount of stochastic delay and packet-loss. This fact strengthens the high reliability of using $PI^\lambda D^\mu$ controllers instead of conventional PID controllers for optimal set-point tracking task over NCS while handling unstable process dynamics.

Table 3:
Tuning results of FOPID controller for FODUP process $P_1$

| Algorithm | $J_{min}$ | $K_p$ | $K_i$ | $K_d$ | $\lambda$ | $\mu$ |
|---|---|---|---|---|---|---|
| DE/rand/1 | 50.28192 | 2.522454 | 1.470881 | 0.182351 | 0.989966 | 0.766836 |
| DE/local-to-best/1 | 50.03256 | 2.549507 | 1.560718 | 0.178657 | 0.989191 | 0.796692 |
| DE/best/1 with jitter | 50.0357 | 2.425323 | 1.365907 | 0.181526 | 0.989938 | 0.828449 |
| DE/rand/1 with per-vector-dither | 50.3835 | 2.444472 | 1.454767 | 0.254753 | 0.98853 | 0.739514 |
| DE/rand/1 with per-generation-dither | 50.27081 | 2.506727 | 1.465598 | 0.150493 | 0.989629 | 0.872172 |
| GA | 50.67115 | 2.253134 | 1.849129 | 0.458813 | 0.986248 | 0.567254 |

Table 4:
Tuning results of FOPID controller for SODUP process $P_2$

| Algorithm | $J_{min}$ | $K_p$ | $K_i$ | $K_d$ | $\lambda$ | $\mu$ |
|---|---|---|---|---|---|---|
| DE/rand/1 | 239.1424 | 0.268791 | 0.445014 | 3.889879 | 0.70663 | 0.438587 |
| DE/local-to-best/1 | 237.1667 | 0.312672 | 0.555823 | 4.465592 | 0.707158 | 0.488474 |
| DE/best/1 with jitter | 236.8556 | 0.401094 | 0.478495 | 4.220372 | 0.734643 | 0.468884 |
| DE/rand/1 with per-vector-dither | 255.1534 | 0.929085 | 0.506063 | 4.678242 | 0.768115 | 0.540872 |
| DE/rand/1 with per-generation-dither | 243.3378 | 0.086403 | 0.552509 | 4.312503 | 0.697915 | 0.444587 |
| GA | 247.2716 | 1.4299 | 0.290376 | 3.60552 | 0.889094 | 0.558855 |

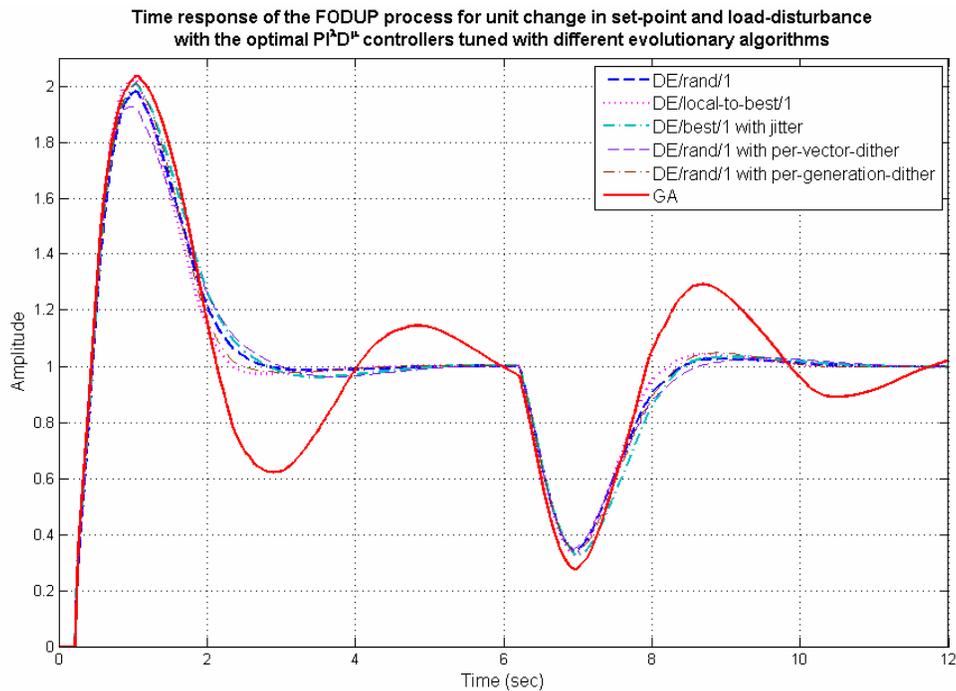

Fig. 12. Time response of $P_1$ over NCS with optimal FOPID controller.

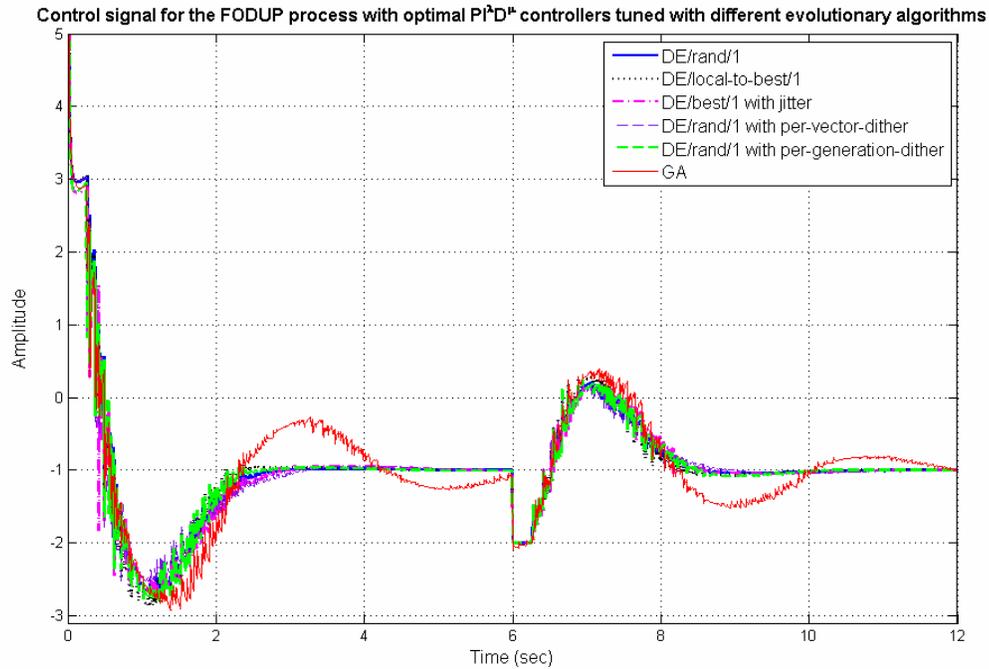

Fig. 13. Control signal of optimal FOPID controller for $P_1$ over NCS.

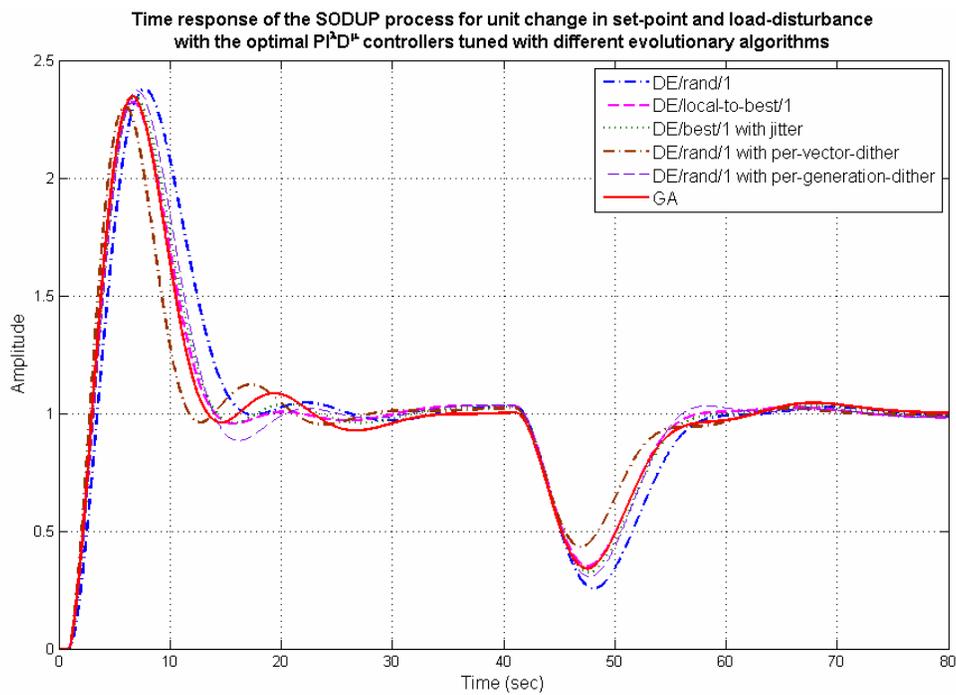

Fig. 14. Time response of $P_2$ over NCS with optimal FOPID controller.

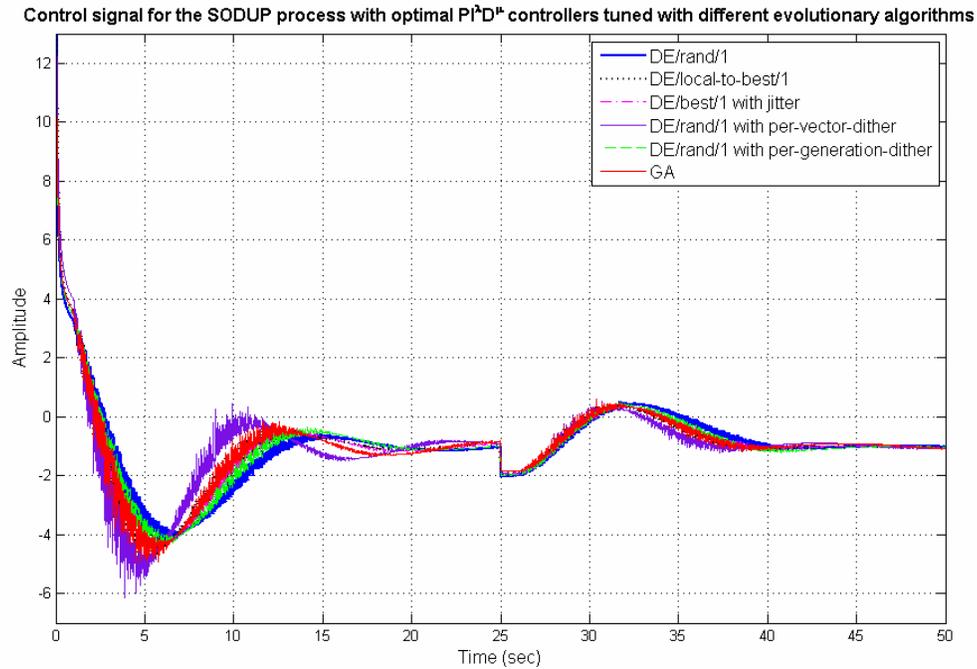

Fig. 15. Control signal of optimal FOPID controller for $P_2$ over NCS.

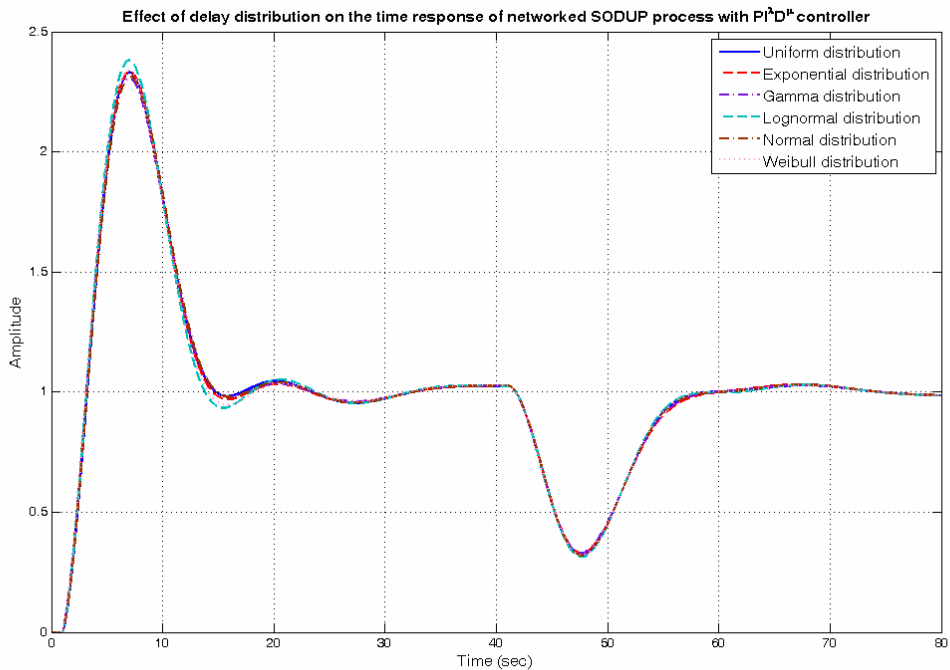

Fig. 16. Effect of change in delay distribution on the time response of the tuned SODUP process.

*4.5. Effect of delay distribution on the tuned networked FOPID control loops*

The tuned FOPID control loop for the SODUP process is tested with different delay distributions to analyze its effectiveness in handling such types of delays though they have not been included in the tuning algorithm. The different parameters of the distributions can be obtained from extensive measurements in a real-time communication channel [64].

In this case, the parameters of the different distributions are so chosen that the delay magnitudes lie between the upper and lower bounds of those considered during the earlier tuning phase. Fig. 16-17 show that the deterioration in control performance for different kinds of delay distribution is negligible and the proposed tuning methodology is capable of handling such variation in delay distribution though they have not been tuned taking these distributions explicitly into account.

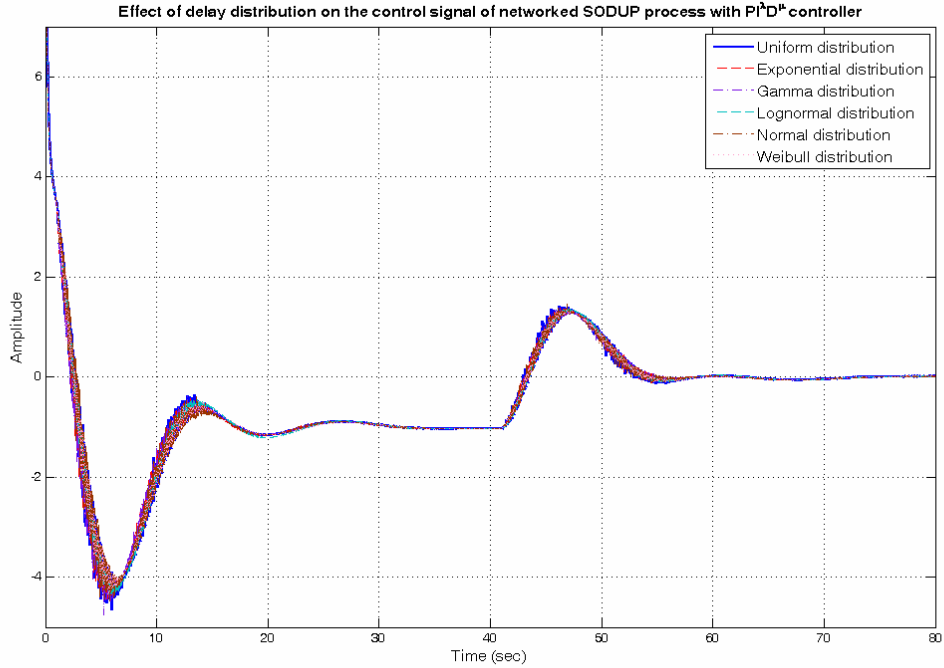

Fig. 17. Effect of change in delay distribution on the control signal of the tuned SODUP process.

### *4.6. Validation of the tuning methodology for lesser complicated FOPTD processes:*

The proposed methodology can be easily validated for commonly encountered FOPTD processes (14) also which are much easier to handle with PID controllers.

$$P_{FOPTD}(s) = \frac{Ke^{-Ls}}{(Ts+1)} \tag{14}$$

The conventional FOPTD class of processes can further be divided in few sub-classes depending on the relative dead-time $\tilde{\tau} = L/(L+T)$. The normalized dead-time closer to zero and one indicates the process to be lag-dominant and delay-dominant respectively. Astrom & Hagglund [65] studied such lag dominated (15) and delay dominated (16) processes which have been tested over a network with drop and variable delays with the proposed tuning technique like that presented in Bhambhani *et al.* [29], [66].

$$P_{FOPTD}^{\tilde{\tau}=0.066}(s) = \frac{1}{(s+1)(0.1s+1)(0.01s+1)(0.001s+1)} \simeq \frac{e^{-0.073s}}{(1.03s+1)} \tag{15}$$

$$P_{FOPTD}^{\tilde{\tau}=0.93}(s) = \frac{e^{-s}}{(0.05s+1)^2} \simeq \frac{e^{-s}}{(0.093s+1)} \tag{16}$$

The GA based PID and FOPID tuning results are shown reported in Table 5 with the corresponding closed loop response and control signals in Fig. 18 for the test lag-dominated and delay-dominated processes. The GA parameters used are same as that in the previous simulations which further proves that a case by case adjustment of tuning algorithm parameters is not necessary. It is seen that both the processes can be better controlled with a FOPID controller in terms of less fluctuation in control signal and set-point tracking and load disturbance rejection performance. Also as indicated in Table 5 the minima is found for a PI or FOPI controller and the derivative action becomes zero or close to zero. This typical behaviour is due to the fact that a derivative action amplifies any randomness in the loop and produces an oscillatory control signal which is harmful from actuator design point of view. However, a simple PI controller is more suitable for FOPTD class of processes as also indicated by Bhambhani *et al.* [66]. However for an open loop unstable process a derivative action is mandatory, since this increases the number of zeros in the complex s-plane, thus attracting the root-locus towards stable negative half-plane. Thus, fractional order derivative action is more suitable for unstable processes over network which is a trade-off between increasing stability by addition of complex zeros and amplification of stochastic behaviour of the data packets in the control loop.

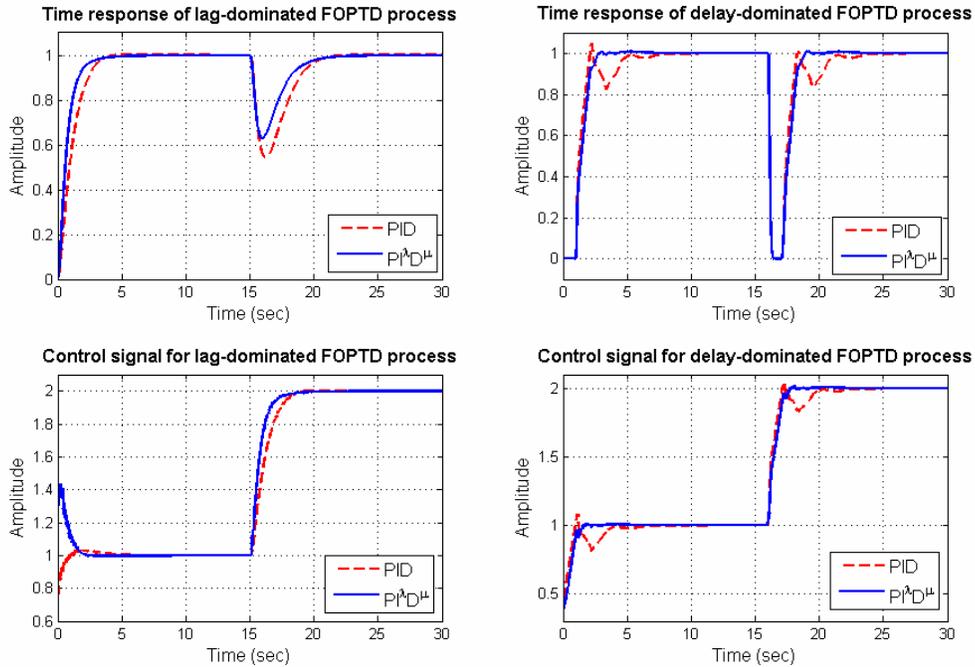

Fig. 18. Tuning performance of lag and delay dominated processes over NCS with the optimally tuned PID and FOPID controllers.

Table 5:
GA based tuning results for FOPTD processes

| Process | Controller Type | $J_{min}$ | Optimal Controller Parameters ||||| 
|---|---|---|---|---|---|---|---|
| | | | $K_p$ | $K_i$ | $K_d$ | $\lambda$ | $\mu$ |
| Lag-dominated | $PI^\lambda D^\mu$ | 41.64 | 1.2543 | 1.2012 | 0 | 0.9900 | 0 |
| | PID | 41.13 | 0.7255 | 0.8477 | 0 | - | - |
| Delay-dominated | $PI^\lambda D^\mu$ | 41.0821 | 0.0617 | 0.6522 | 0.2267 | 0.9958 | 0.1205 |
| | PID | 41.3092 | 0.4375 | 0.5755 | 0 | - | - |

## 5. Conclusion:

The FOPID and PID controllers are tuned for time domain optimality criterion with GA and few variants of DE to handle the adverse effects of stochastic variation of network delay and packet dropouts. The delays have been considered to be greater than one sampling time unlike [30], hence there is a probability of out of order packet arrival which is handled by a time stamped buffer allowing only the most recent packets to pass. Among the different evolutionary algorithms the DE variants perform better at controller tuning than GA. The paper also shows the adverse effect of controller tuning with lumped approach of the stochastic network delay unlike [12]. The simulation results show that the $PI^\lambda D^\mu$ controllers are more suitable for NCS applications than their corresponding PID counterparts, as they not only give better time domain performance under uncertain network conditions, but also produces significantly smaller control signal which would require a smaller actuator size and reduce the risk of actuator saturation. For highly unstable plants like $P_2$ no stabilizing PID controller can be found which can handle the delays and packet dropouts of the network simultaneously. However the same could be stabilized with good time domain performance using a FOPID controller. This further establishes the suitability of FOPID controllers in NCS applications. The tuning methodology is validated for other less critical processes like delay and lag dominated FOPTD. It is also shown that the parameters of the optimization algorithms for controller tuning are robust enough to work on a wide variety of processes and need not be changed case by case. The tuned controller parameters are shown to work well for other type of network delay distributions. Future work would involve real-time implementation of these controllers over the network to verify the design methodology.

**Acknowledgement**

This work has been supported by the Board of Research in Nuclear Sciences (BRNS) of the Department of Atomic Energy (DAE), India, sanction no. 2009/36/62-BRNS, dated November 2009.